# Absorptive loss and band non-parabolicity as a physical origin of large nonlinearity in epsilon-near-zero materials

Ray Secondo[1], Jacob Khurgin[2], and Nathaniel Kinsey[1,*]

[1]Dept. of Electrical and Computer Engineering, Virginia Commonwealth University, Richmond, VA 23284, USA

[2]Dept. of Electrical and Computer Engineering, Johns Hopkins University, Baltimore, MD 21218, USA

[*]nkinsey@vcu.edu

**Abstract:** For decades, nonlinear optics has been used to control the frequency and propagation of light in unique ways enabling a wide range of applications such as ultrafast lasing, sub-wavelength imaging, and novel sensing methods. Through this, a key thread of research in the field has always been the development of new and improved nonlinear materials to empower these applications. Recently, epsilon-near-zero (ENZ) materials have emerged as a potential platform to enhanced nonlinear interactions, bolstered in large part due to the extreme refractive index tuning ($\Delta n \sim 0.1 - 1$) of sub-micron thick films that has been demonstrated in literature. Despite this experimental success, the theory has lagged and is needed to guide future experimental efforts. Here, we construct a theoretical framework for the intensity-dependent refractive index of the most popular ENZ materials, heavily doped semiconductors. We demonstrate that the nonlinearity when excited below bandgap, is due to the modification of the effective mass of the electron sea which produces a shift in the plasma frequency. We discuss trends and trade-offs in the optimization of excitation conditions and material choice (such material loss, band structure, and index dispersion), and provide a figure of merit through which the performance of future materials may be evaluated. By illuminating the framework of the nonlinearity, we hope to propel future applications in this growing field.

## 1. Introduction

Nonlinear optics studies the reaction of materials to intense light and has long been an avenue of interest for controlling the flow of light [1,2]. These effects can be utilized to dynamically control the phase, amplitude and frequency of light within a wide range of materials such as crystals (LiNbO₃, KTP) [3], semiconductors (GaAs) [3], and polymers (PTS, MAP) [4], although the weak response typically requires the use of long propagation lengths or large optical intensities. Consequently, a significant effort has been invested to maximize the nonlinearity through avenues such as bulk material optimization [3–6] and nanostructuring [7–10]. Among these, recent success has been found working with epsilon-near-zero (ENZ) materials – media that exhibit a spectral range where $|\operatorname{Re}\{\varepsilon(\omega)\}| < 1$ [11–21]. Physically, this condition may be satisfied in bulk materials near resonances or through free carriers, as well as in

nanostructured materials as an effective property by mixing both metals and dielectrics. [13] For low-loss ENZ materials, the real index is also less than unity – a condition termed near-zero index (NZI) – and such materials have demonstrated an enhancement of nonlinear processes including the intensity-dependent refractive index (IDRI) [20,22–25] and frequency conversion [14,26–28] through electric field confinement and slow light effects [29,30], thereby opening a breadth of applications in light manipulation [25]. It should be noted that an effective NZI condition may also be achieved in all-dielectric nanostructures with $\varepsilon > 1$, however, we do not consider this case here as the nonlinearities in this structure are the same as in the constituent dielectrics.

A prominent subset of ENZ materials is the transparent conducting oxides (TCOs) [31–34], such as indium tin oxide (ITO) or aluminum-doped zinc oxide (AZO). These materials provide a carrier concentration of up to $10^{20} – 10^{21}$ ($cm^{-3}$) due to the high fraction of donor atoms $In^{3+}$ in ITO or $Al^{3+}$ in AZO, and produce an ENZ region in the telecommunication spectrum ($1.3 – 1.5$ ($\mu m$)). Such materials have exhibited refractive index modulation that is comparable to the linear index ($\Delta n \approx 0.2 - 1$) [22–24,35,36] while being ultrathin, industrially friendly [37], and CMOS compatible [33]. As a result, index modulation in NZI materials is of interest for potential applications in all-optical switching.

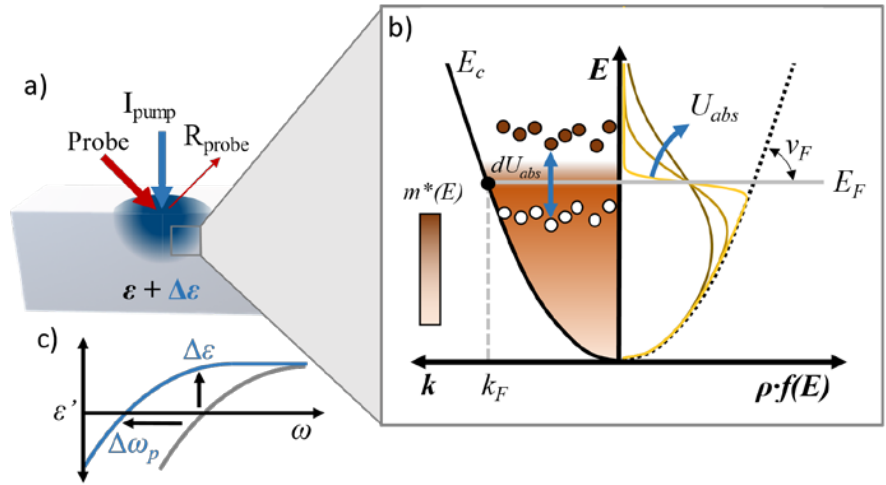

Fig. 1: a) Schematic of the intraband nonlinearity in ENZ materials, where the reflectivity (permittivity) of the material is changed through the application of a pump beam. b) The change in permittivity occurs due to a modification of the effective mass of the electron sea as the absorbed pump energy elevates electrons to higher energy, higher mass states. c) The resulting change in effective mass red-shifts the plasma frequency of the material, producing the modulation of permittivity at a fixed frequency.

The continuation of these studies necessitates effective modeling to guide experimental and material synthesis efforts. Through previous works, the IDRI of NZI materials has been modeled using the Kerr-process description [24]. While this approach has been shown to provide a reasonable quantitative description, it implicitly assumes a polarization driven effect while the nonlinearity has been experimentally shown to be due to free-carrier effects [35,38]. Therefore, to further explore the mechanisms of the nonlinearity, predict ideal regimes of operation, and optimize materials, a more physically accurate model is worth investigation.

This work aims to develop a deterministic (fit-parameter free) physical model to describe the large nonlinearities in Drude NZI materials, such as TCOs, to gain physical insight into the underlying mechanisms as well as predict the optimal material and experimental requirements. Through this, we emphasize that the driving force of the effect is due to the non-parabolic dispersion of the conduction band and the resulting effect on the average effective mass under intraband excitation.

## 2. Nonparabolicity of the Band as the Cause of Nonlinearity

In ENZ materials, the intensity-dependent refractive index is generally induced through an intraband absorption process whose effects are schematically shown in Fig. 1 (although other methods are possible such as interband excitation). In this case, energy from the pump is absorbed by free carriers ($U_{abs}$), thereby raising their kinetic energy and the electron temperature $T_e$, Fig. 1b depicted by the smeared occupied state distribution $\rho(E) \cdot f(E)$, where $\rho(E)$ is the density of states and $f(E)$ is the Fermi-Dirac probability function. As the distribution of carriers shifts to higher energy states, the average effective mass of the electron sea tends to increase since the band is non-parabolic (illustrated by the color gradient for free-carriers in the conduction band). This, in turn, causes a decrease in the plasma frequency and a corresponding increase in dielectric constant (Fig. 1c). The process relaxes as the kinetic energy of the hot electrons is transferred to the lattice (i.e. through acoustic phonons) with the characteristic electron-lattice relaxation time $\tau_{el}$. From this picture, it is clear that several parameters play a key role in the strength of the process: 1) the absorbed energy, 2) the band nonparabolicity, and 3) the dispersion in the index, which is quite different from typical polarization-driven nonlinearities.

To provide more detail into the IDRI in ENZ materials, we begin by defining the momentum-dependent mass of the electron. This can be determined by considering how the momentum of electrons at the Fermi surface change due to an applied electric field $\boldsymbol{F}$. Here, the field shifts the Fermi-surface in $k$-space (see Fig. 2a), producing a change in momentum that for some state $k$ is given by:

$$\delta \boldsymbol{k} = -e\boldsymbol{F} / \hbar.$$  (1)

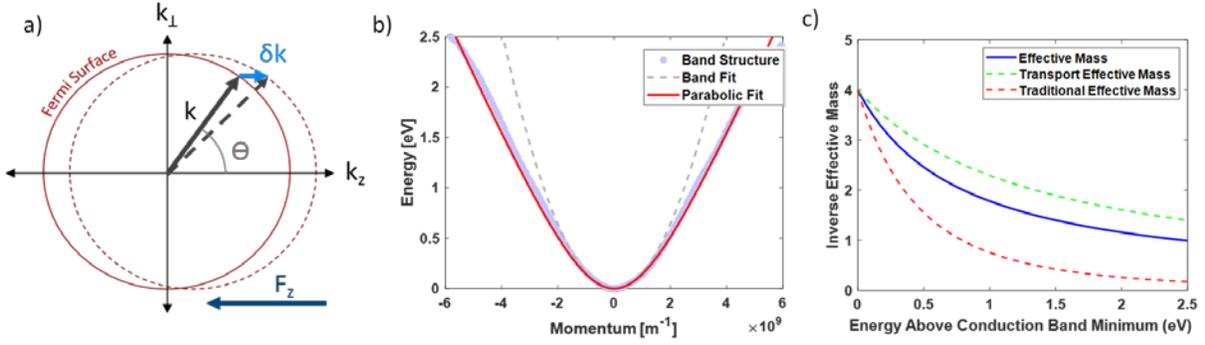

Fig. 2: a) Schematic of the change in momentum for electrons near the Fermi-level under an applied field. b) Energy band dispersion adapted from Ref. [39] and the hyperbolic fit used for calculations compared to a parabolic fit. c) Comparison of the inverse effective mass of electrons for doped zinc oxide films at room temperature versus energy into the conduction band.

For a field applied along the z-direction, the velocity associated with a particular momentum state becomes:

$$v_z = \cos \theta = v_F \frac{k_z}{\sqrt{k_z^2 + k_\perp^2}}$$  (2)

where $\theta$ is the angle between state $\boldsymbol{k}$ and the axis of the applied field, in this case, the z-axis. We can then introduce the effective mass of the electron, described by the rate of change of the velocity versus the rate of change of the momentum, $dv/dk$. Thus, when we take the full-derivative with respect to $k_z$ we obtain:

$$\frac{dv_z}{dk_z} = \frac{dv(k)}{dk_z} \cos \theta - v(k) \sin \theta \frac{d\theta}{dk_z} = \frac{dv(k)}{dk} \cos^2 \theta + \frac{v(k)}{k} \sin^2 \theta$$  (3)

Averaging over the polar angle, $\left\langle \sin^2 \theta \right\rangle_\theta = 2/3$ and $\left\langle \cos^2 \theta \right\rangle_\theta = 1/3$, and taking into account that the field is harmonic, we obtain:

$$\delta v_z = -\frac{e}{i\omega\hbar} \left[ \frac{2}{3} \frac{v(k)}{k} + \frac{1}{3} \frac{dv(k)}{dk} \right] F = -\frac{e}{i\omega m^*(k)} F$$  (4)

where

$$m^*(k) = \hbar \left[ \frac{2}{3} \frac{v(k)}{k} + \frac{1}{3} \frac{dv(k)}{dk} \right]^{-1} \qquad (5)$$

is the momentum dependent effective mass. Note that this definition is different from the more conventional definition of effective mass as the radius of curvature in $k$-space, $m^*(k) = \hbar^2 (\partial^2 E / \partial k^2)^{-1}$, and the less common transport effective mass [40,41] $m^*(k) = \hbar^2 k (\partial E / \partial k)^{-1}$. In fact, the above formula is a weighted average of the two, simplifying to the first description in the case of a perfectly parabolic band, and to the latter in the case of a perfectly linear band. For a non-parabolic ZnO band, as shown in Fig. 2b (energy band data from ref. [39]), the definitions of the effective mass are compared in Fig. 2c, where the difference between the definitions is apparent.

Given a general description for the effective mass of a state, we can then describe the optical response of the material under an applied field of arbitrary frequency by summing the induced displacements for all electrons in the gas, $z(t) = -i\omega^{-1} v_z(t)$, as:

$$\varepsilon(\omega) = \varepsilon_\infty - \frac{q^2}{\varepsilon_0 (\omega^2 + i\omega\Gamma)} \frac{1}{V} \sum_{l=1}^{n_{car}} \frac{1}{m^*(k_l)} = \varepsilon_\infty - \frac{Nq^2}{\varepsilon_0 (\omega^2 + i\omega\Gamma) m^*_{avg}} \qquad (6)$$

which produces a generalized Drude formula for non-parabolic bands, where $\Gamma$ is the scattering rate, $l$ is an occupied state, and $n_{car}$ is the number of carriers in the volume $V$. The average effective mass of the electron gas, $m^*_{avg}$, is then:

$$m^*_{avg} = \left[ \frac{1}{NV} \sum_{l=1}^{n_{car}} \frac{1}{m^*(k_l)} \right]^{-1} \qquad (7)$$

or for a continuous energy band as:

$$m^*_{avg} = \frac{1}{N} \int \frac{f(E, \mu, T_e)}{m^*(E)} \rho(E) dE \qquad (8)$$

where $N$ is the carrier density, $\rho(E)$ is the density of states, and $f(E, \mu, T_e)$ is the Fermi-Dirac distribution:

$$f(E, \mu, T_e) = \left[ \exp\left( \frac{E - \mu}{k_B T_e} \right) + 1 \right]^{-1} \qquad (9)$$

with chemical potential $\mu$ and electron temperature $T_e$.

The effective mass of the electron sea is thus given by the geometric average of the occupied states, whose distribution in energy (e.g. described by the Fermi-Dirac probability) can be modified by absorbing the energy of an optical pump through the intraband absorption process. In this case, the change in permittivity can be found by differentiating Eq. 6 with respect to the incident pump intensity, $I$, where we find it is proportional to the change in effective mass:

$$\delta\varepsilon(I) = +\frac{Nq^2}{\varepsilon_0 m^*_{avg}(\omega^2 + i\omega\Gamma)}\frac{\delta m^*_{avg}(I)}{m^*_{avg}} \quad . \tag{10}$$

When evaluated within the ENZ region at the screened plasma frequency of the material where $\varepsilon'(\omega_p) = 0$ and $\varepsilon_\infty \approx Nq^2/\left(\varepsilon_0 m^*_{avg}\omega_p^2\right)$, the expression simplifies to:

$$\delta\varepsilon\Big|_{\omega_p} \approx \varepsilon_\infty \frac{\delta m^*_{avg}(I)}{m^*_{avg}} \quad . \tag{11}$$

Ultimately, this expression can be interpreted as a change of the screened plasma frequency, resulting in a red-shift of the entire dispersion curve as shown in Fig.1c.

## 3. Estimating the Strength of the Nonlinearity

We can provide an estimate for the strength and dependence of the nonlinearity by taking a few assumptions in order to obtain an analytical expression for the nonlinear coefficients in the ENZ region. For AZO with an ENZ point near the telecom wavelengths, the carrier density must be approximately $N \sim 0.8 \times 10^{21}(cm^{-3})$ [42]. In this case, the Fermi level resides well above the bottom of the conduction band (by as much as $1\ (eV)$) [43]. As a result, we may assume that the average effective mass of the electron gas is approximately equal to the effective mass at the Fermi Level, $m^*_{avg} \approx m^*(k_F)$, and that the velocity is nearly saturated at the Fermi level, i.e. $dv(k_F)/dk << v(k_F)/k_F$ (see Fig. 2b). This simplification allows us to directly write the permittivity of the material using Eq. 6 and provides a good estimate of the nonlinear effect without making any assumption on the distribution of hot electrons. Furthermore, it does not require the introduction of an electron temperature! A similar assumption can be used for many other heavily doped semiconductors where the band dispersion in quasi-linear at the Fermi level.

Now, because the number of carriers in the band must be conserved during the intraband process, the change in permittivity is then proportional to the overall variation in the average effective mass due to the absorption of the pump photons. When electrons absorb a pump photon, their energy changes by some amount $\delta E_i$, which for electrons near the Fermi level produces a change of momentum for the electron of $\delta k_i = \delta E_i / \hbar v_F$, where $v_F$ is the Fermi velocity (see Fig. 1b). Given our momentum dependent effective mass, the change in the effective mass due to the intraband absorption process is:

$$\delta\left(\frac{1}{m^*}\right)_i = \left(\frac{\partial}{\partial k}\frac{1}{m^*(k)}\right)_i \delta k = -\frac{1}{m^*(k_F)}\frac{\delta k_i}{k_F} = -\frac{1}{m^*(k_F)}\frac{\delta E_i}{\hbar v_F k_F} \quad . \tag{12}$$

Although $\delta E_i$, and therefore $\delta k_i$, is different for each electron, the sum of the energy change over all of the electrons within a given volume is equal to $V^{-1}\sum_i \delta E_i = \delta U_{abs}$, where $\delta U_{abs}$ is the absorbed energy density of the pump beam. We can then find the total change to the permittivity by summing over the small modifications in the effective mass induced by the excitation of hot electrons which while neglecting loss becomes:

$$\delta\varepsilon \approx -\frac{q^2}{\varepsilon_0\omega^2}\frac{1}{V}\sum_{\delta k}\delta\left(\frac{1}{m^*}\right)_i = \frac{q^2}{\varepsilon_0 m^*(k_F)\omega^2}\frac{\delta U_{abs}}{\hbar v_F k_F} \tag{13}$$

and when using the relation $\varepsilon_\infty = Nq^2 / \left(\varepsilon_0 m^*_{avg}\omega_p^2\right)$ reduces to:

$$\delta\varepsilon(\omega_p) = \varepsilon_\infty \frac{\delta U_{abs}}{N\hbar v_F k_F} \tag{14}$$

at the screened plasma frequency. In general, the average absorbed energy density over the sample thickness $d$ can be written as:

$$\delta U_{abs} = I_i(1 - e^{-\alpha d})\tau / d \tag{15}$$

where $I_i$ is the pump intensity just inside the sample and related to the external intensity $I$ as $(1 - R)I$ where $R$ is the reflectivity of the sample at $\omega_{pump}$, and $\tau$ is the time over which the energy of the pulse gets accumulated in hot carriers. For CW illumination or long pulses, this time is $\tau_{el}$, the electron-lattice relaxation time of the system, while for shorter pulses with $\tau_p < \tau_{el}$, the accumulation time is $\tau_p$ itself (see Appendix A). Under a thin sample approximation $d \ll \alpha^{-1}$ and considering the pump electric field just inside the material $F_i$, this becomes:

$$\delta U_{abs} = \frac{\alpha_o n' n_g \tau}{2 \eta_0} F_i^2 \tag{16}$$

where $n'$ is the real part of the refractive index at $\omega_{pump}$, $\eta_0$ is the free space impedance, and $\alpha_o$ is the nominal non-enhanced absorption coefficient at $\omega_{pump}$. However, slow light effects caused by the low refractive index, or large group index $n_g$, induce an enhanced absorption $\alpha \approx \alpha_o / n' = n_g \alpha_o$ [30].

Using Eq. 13, Eq. 15, and the definition of the nonlinear susceptibility, $\delta \varepsilon = F_i^2 \chi_{eff}^{(3)}$, the real part of the effective third-order susceptibility can be estimated as:

$$\chi_{eff}^{(3)'}(\omega_p) = \varepsilon_\infty \frac{\alpha_o n' n_g \tau}{2 \eta_0 N \hbar v_F k_F} = \varepsilon_\infty \frac{\alpha n' \tau}{2 \eta_0 N E_F} \tag{17}$$

from which, the imaginary part of the susceptibility can found as:

$$\chi_{eff}^{(3)''}(\omega_p) = -\chi_{eff}^{(3)'}(\omega_p) \frac{\Gamma(\omega_p)}{\omega_p} = -\varepsilon_\infty \frac{\alpha n' \Gamma(\omega_p) \tau}{2 \omega_p \eta_0 N E_F} \tag{18}$$

Now we can also calculate the IDRI coefficient and nonlinear absorption as in ref. [1] :

$$n_2(\omega_p) = \varepsilon_\infty \frac{\alpha_o n_g \tau}{2 n'(\omega_p) N E_F} = \varepsilon_\infty \frac{\alpha \tau}{2 n'(\omega_p) N E_F} \tag{19}$$

$$\alpha_2(\omega_p) \approx 2 \eta_o \frac{\omega}{c} \frac{\chi_{eff}^{(3)''}}{n'^2(\omega_p)} = -\varepsilon_\infty \frac{\alpha n' \Gamma(\omega_p) \tau}{c n'^2(\omega_p) N E_F} \tag{20}$$

where $n'(\omega_p)$ is the real part of the index at the screened plasma frequency and $\Gamma$ is the Drude scattering rate, determined through the mobility as $\Gamma = q / \left( \mu_e m_{avg}^* (I = 0) \right)$. For a heaviliy doped semiconductor with ENZ wavelength

$\lambda \approx 1330\ (nm)$, we can approximate $n_2$ with estimations of $\varepsilon_\infty \approx 3.5$, $\alpha$ on the order of $1 \times 10^5 (m^{-1})$, $\tau = \tau_p = 100\ (fs)$, $n' \approx 0.3$, carrier density $N \approx 1 \times 10^{21} (cm^{-3})$ and $E_f \approx 4.5 (eV)$ (1 eV beyond the band gap). The intensity-dependent index is then estimated as $n_2 \sim 8 \times 10^{-17} (m^2/W)$ compared to a value of $3 \times 10^{-17} (m^2/W)$ for AZO reported in literature [24].

## 4. Predicting Experiments

To move beyond the assumptions inherent in Eq. 19, we retain the material loss and incorporate the complete dispersion of the effective mass which can be determined from a given material band structure. In this case, we directly determine the average effective mass of the electron sea using Eq. 8. Therefore, calculating the nonlinearity reduces to determining the distribution of the electrons within the band, and thus, the modification of $m^*_{avg}$.

For a given pump intensity, the distribution of carriers, described through the Fermi-Dirac probability, can be found using conservation identities of charge and energy. The first states that the number of carriers residing in the band must remain constant under intraband excitation:

$$\int f(E, \mu(I), T_e(I))\rho(E)dE = N\ .$$ (21)

The second states that the total change in energy of the electron gas must be equal to the absorbed energy such that:

$$\int f(E, \mu(I), T_e(I))E\rho(E)dE - \int f(E, \mu_0, T_0)E\rho(E)dE = \delta U_{abs} = AI\tau/d$$ (22)

where $\mu_0$ is initial chemical potential before pumping, found for $T_e = T_0$, and $\tau$ is either $\tau_{el}$ or $\tau_p$, chosen as described for Eq. 15.

Using these constraints, the chemical potential and electron temperature can be uniquely defined for a given pump intensity $I$, and the average effective mass can then be found then as $\dfrac{1}{m^*_{avg}(I)} = \int \dfrac{f(E, \mu(I), T_e(I))E\rho(E)}{m^*(E)}dE$, leading to the intensity-dependent permittivity is:

$$\varepsilon(\omega, I) = \varepsilon_\infty - \frac{Nq^2}{\varepsilon_0 m^*_{avg}(I)}\frac{1}{\omega^2 + i\omega\Gamma}\ .$$ (23)

In general, the Drude scattering rate $\Gamma$ will also be dependent upon the incident pump intensity (see Appendix A), but in the case of heavily doped TCOs such as AZO and ITO, which are known to contain a large density of defects [43–45], $\Gamma$ is assumed to remain constant, essentially being limited by the defect and ionized donor scattering rate. Furthermore, a variation in $\Gamma$ produces a change to the real permittivity of $\delta\varepsilon' \sim 2\varepsilon_\infty \Gamma(\delta\Gamma/\omega^2)$, which is negligibly small compared to the modified effective mass. From Eq. 23, we can then determine the change in permittivity and index as a difference between the properties of the steady-state and the pumped material.

Examining the limits of the nonlinearity, we find that the max variation in the refractive index is achieved when the change in effective mass is also maximized, $\delta n_{max} = \sqrt{\varepsilon_\infty} - n_0$, essentially resulting in complete suppression of the Drude contribution. In this sense, saturation is inherently built into the model, not requiring the introduction of higher-order nonlinear susceptibilities to explain. This occurs naturally because $\delta\varepsilon$ is inversely proportional to the change in the average effective mass, though material breakdown and other non-ideal effects may occur at high fluences. Furthermore, in contrast to polarization driven nonlinearities this change occurs over an "integration time" of $\tau$ leading to 1) an increased nonlinear effect, since the pump energy feeding the nonlinearity continues to build throughout the interaction and 2) a sub-picosecond rise and fall time. Although not instantaneous, this response time is sufficient for numerous potential ultrafast applications and enables the use of the larger nonlinear phenomena while maintaining a strong overlap to many modern ultrafast laser systems with pulse-widths on the order of 10's to 100's $(fs)$.

To verify the model, experimental data from literature studying IDRI in AZO films at ENZ is used [24]. In this case, the sample was interrogated under non-degenerate conditions where $\lambda_{pump} = 785 \ (nm)$ and $\lambda_{probe}$ is varied from $1150 - 1550 \ (nm)$. The change in the index $\delta n$ was calculated using an inverse transfer matrix method from the measured change in reflection and transmission of the film (see [24] for additional details). To approximate the measured films, the conduction band dispersion of AZO is taken as a hyperbola fit to the undoped ZnO band structure [39] (see Appendix B). The permittivity is modeled using Eq. 23, and the absorption of the film is calculated through the transfer matrix method. Eq. 22 is used with an exponential relaxation model of characteristic time $\tau_{el} = 170 \ (fs)$ [20] to calculate the carrier distribution at increased temperatures (see Appendix C). This correction is used because the pump pulse width ($\tau_p = 100 \ (fs)$) is comparable to the electron-phonon relaxation rate such that the peak temperature of the electron gas is reduced by the competing absorption and phonon scattering processes.

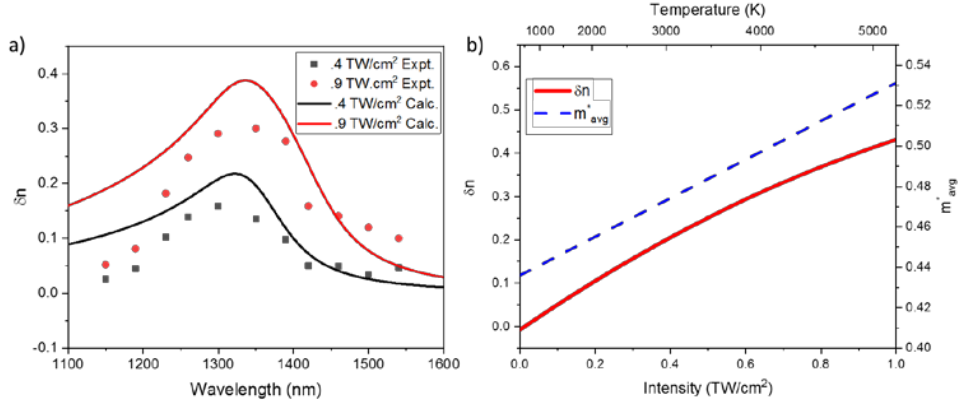

Fig. 3: a) The change in refractive index due to a $0.4$ $(TW/cm^2)$ (black) and $0.9$ $(TW/cm^2)$ (red) pump at 780 nm. Using a relaxation rate of $\tau = 170$ $(fs)$ [20] on a Gaussian pulse shape with maximum pump with 100 fs full-width half-maximum. A close fit between experimental [24] and theoretical responses is obtained through a deterministic model. b) The peak $\delta n$ and average effective mass of carriers versus the applied pump intensity for the AZO sample.

Fig. 3a demonstrates that the model can accurately predict the index change for the AZO film in both magnitude and spectrum near the ENZ point of $1330$ $(nm)$ for multiple fluence levels without free fitting parameters. The fit deviates at short wavelengths due to the discrepancy of the permittivity of the approximated and measured permittivity of the sample [24]. Breaking this down we can extract the average effective mass induced in the AZO film and peak change in the index within the ENZ region, Fig. 3b, as a function of the pumping intensity. Here, the average effective mass is shown to vary linearly with the pump intensity, producing an increase of ~20% for 1 $(TW/cm^2)$. However, the peak change of the refractive index in the ENZ region is observed to be nonlinear with intensity. This is expected since the change in the index is inversely proportional to the change in $m_{avg}^*$. However, this demonstrates that a single $n_2$ value is a limited description of the nonlinearity in ENZ materials.

## 5. Discussion

Given this description, one can then consider what nonlinear material (e.g. doped ZnO, GaN, SnO) and what material properties (e.g. $\varepsilon_\infty$, $\mu_e$, $\Gamma$) would produce the most efficient response. To answer, we can return to the three performance metrics for the intraband effect, 1) absorption, 2) band non-parabolicity, and 3) index dispersion, to can construct a material figure of merit (FOM).

First, we consider the material's ability to utilize the incident pump energy, which is quantified by the absorption of the material $A_{pump} \approx (1-R)\alpha d$, as the beginning of our FOM. Clearly from Eq. 13 and Eq. 22, if a material can absorb more of the incident pump energy, a larger modification to the electron distribution within the conduction band will occur.

Next, we consider how the average effective mass depends upon the distribution of the electrons within the band. This is quantified by the change in the average effective mass for a fixed amount of absorbed energy which can be approximated by the slope of the effective mass at the Fermi level, $dm^*(k_F)/dE$. Noting that materials with a larger initial effective mass require an increased modulation to produce the same shift in the permittivity, we include the change in the normalized effective mass $(m^*_{avg})^{-1}[dm^*(k_F)/dE]$ into our FOM. The range of $dm^*(k_F)/dE$ is generally determined by the degree of nonparabolicity of the conduction band, and as a rule of thumb, is inversely proportional to the energy gap.

Finally, we consider the sensitivity of the material to a change in the effective mass, $dn_{probe}/dm^*_{avg}$. Since the modification of the average effective mass effectively shifts the plasma frequency of the material, for index modulation $dn_{probe}/dm^*_{avg} \propto dn_{probe}/d\omega$, such that materials with an increased index dispersion at the probe wavelength (typically in the ENZ region) will be optimal. This can be achieved in two ways 1) through a reduced loss, and 2) a larger background permittivity $\varepsilon_\infty$. For a material of choice, the first point represents a competing optimization case as $dn_{probe}/d\omega$ will be maximized at ENZ in a low loss medium while $A_{pump}$ is reduced, irrespective of the pump frequency. This suggests that for any material, there is an optimum loss value that balances the necessary absorption with reflection, in contrast to a Kerr nonlinearity where reduced loss is always favored. The second point can be illustrated by considering a lossless material where $\delta\varepsilon = \varepsilon_\infty \, \delta m^*/m^*$ (Eq. 11), that allows for a simplification of $dn_{probe}/dm^* \approx (\varepsilon_\infty/m^*_{avg})^{1/2}$. Ultimately, an increased $\varepsilon_\infty$ requires a larger contribution from the free-electrons to produce an ENZ region at a fixed wavelength, which results in a steeper index dispersion. However, this is generally accomplished by increasing the carrier density of the material. As a result, more electrons must undergo an interband transition (i.e. more pump energy must be absorbed) to produce the same shift in permittivity, leading to a factor $1/N$.

It is important to note that the nonlinearity also depends on accumulation time $\tau$ (e.g. through the absorbed energy as seen in Eq. 15), but the bandwidth of the nonlinear effect is inversely proportional to this time. As a result, we can define a FOM by combining these quantities which form an $n_2$-strength-bandwidth product that can be used as an intrinsic material FOM for intraband nonlinearities:

$$FOM = A_{pump}\left[\frac{1}{m^*_{avg}}\frac{dm^*}{dE}\right]\left[\frac{1}{N}\frac{dn_{probe}}{dm^*_{avg}}\right] \quad (24)$$

To explore this, the band structures of indium tin oxide (ITO), cadmium oxide (CdO), gallium nitride (GaN), and gallium oxide (Ga$_2$O$_3$) were approximated from literature and the FOM for each material was calculated versus mobility. The results are summarized in Fig. 4 while the values used for the calculation are shown in Table 1 where the listed mobility ranges correspond to published values for each material. The films are considered to be under the same experimental conditions as in [24]. As a reference, the peak FOM of each material within the range of mobility values reported in literature is also included. The values of $A_{pump}$ are calculated through the transfer matrix method and $dn_{probe}\big/dm^*_{avg}$ is calculated through a Drude-Lorentz model where the carrier density in the materials has been modified to achieve an ENZ at 1325 $(nm)$ for all materials (see Appendix D for details).

The first general observation that can be made from this comparison is that vastly different conductive oxides and nitrides have a FOM that differs by just more than a factor of three, as can be expected based on the values as reported in Table 1. The second observation is that based on Fig. 4 the peak FOM is achieved for moderate mobility films in the range of $\mu_e \approx 15$ $(cm^2 / V - s)$ which for ZnO yields $\varepsilon''(\omega_{probe}) = 0.6$. This illustrates the important role of loss in the optimization as high mobility films suffer from decreased absorption while low mobility films suffer from a shallow index dispersion. Moreover, it is evident that doped zinc oxide (AZO, GZO, IZO) outperforms the other materials, including slightly outperforming ITO, due to the moderately low $m_{avg}^*$ and the lowest $N$ required for near-infrared ENZ of all the considered materials, see Table 1. Although a similar result has been noted in literature [12], an exact comparison is strongly dependent upon the material growth conditions as the effective mass varies with factors such as defects and grain size [41] while the band structure varies with stress/strain and doping density among others.

Table 1. Approximate comparison of transparent semiconductors for intensity-dependent refractive index calculations. CdO, GaN, and Ga$_2$O$_3$ mobilities are for lower carrier concentrations ($< 0.5 \times 10^{20} (cm^{-3})$). N and $m^*_{avg}$ are calculated for all materials to generate an ENZ wavelength of **1325 $(nm)$** (See Appendix D).

| Material | $\mu_e (cm^2 / V - s)$ (Highly Doped) | $\varepsilon_\infty$ | N ($\times 10^{20} cm^{-3}$) | $m^*_{avg}$ | $\dfrac{dm^*(k_F)}{dE}$ $(eV^{-1})$ | Maximum Figure of Merit $(m^3/eV)$ |
|---|---|---|---|---|---|---|
| ZnO [39,41,46,47] | 10 [46]  -  40 [47] | 3.3 | 9.19 | $0.44 m_o$ | 0.36 [39] | 69.8 |
| ITO [44,48] | 20  -  35 [44] | 3.66 | 11.3 | $0.49\ m_o$ | 0.36 [48] | 56.8 |
| Ga$_2$O$_3$ [49–51] | 0.1 [49]  -  25 [50] | 3.57 | 11.8 | $0.52\ m_o$ | 0.40 [51] | 52.9 |
| GaN [52–54] | >100 [52,53] | 5.3 | 13.4 | $0.40\ m_o$ | 0.23 [54] | 35.5 |
| CdO [55–57] | 80 [55]  -  >200 [55,56] | 5.5 | 18.9 | $0.54\ m_o$ | 0.37 [57] | 21.9 |

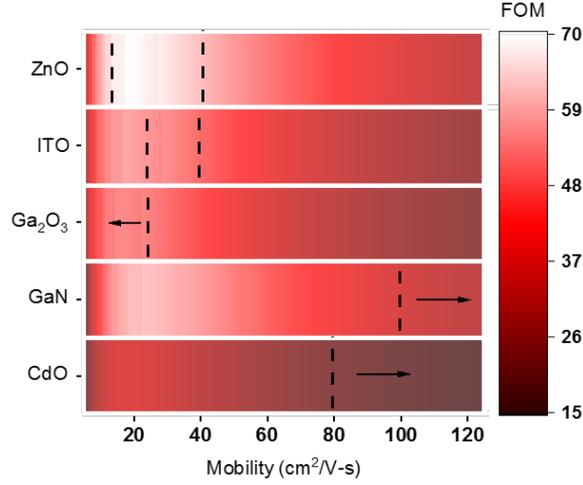

Fig. 4 Figure of merit for the selected materials as a function of optical mobility. Grey dashed lines represent the range of mobilities achieved by films published in literature. If only a single line, an arrow indicates the range of mobilities.

Beyond these more explored films, Ga₂O₃ shows the highest FOM within the range of published mobilities; however, it has not been doped into the range necessary for consideration in the telecommunications window and the performance may vary with heavy doping. Yet, Ga₂O₃ is a uniaxial crystal which may prove useful in exploring nonlinear phenomena that manipulate optical polarization [58]. The performance of CdO and GaN suffers due to their high mobilities which reduce their ability to use the pump energy supplied. However, if the materials are allowed to take any mobility (i.e. considering values outside of the grey dashed regions in Fig. 4), GaN achieves a maximum FOM of 61.0 for a mobility of $\mu_e \approx 23$ $(cm^2 / V - s)$, while CdO finds its maximum near 36.8 for a similar mobility

For lower mobilities, GaN shows a high FOM due to the combination of a large $\varepsilon_\infty$ and low $dm^*(k_F)/dE$ that allows for $N$ and $m^*_{avg}$ to stay comparably low while achieving ENZ in the near-infrared.

## 6. Conclusions

Through analysis of free carrier dynamics, a model has been derived to explain and fit the large nonlinear response of Drude ENZ materials, such as TCOs. We demonstrate that the nonlinearity is due to free-carrier absorption and the subsequent modification of the average electron effective mass through a smeared population in a non-parabolic conduction band. Specifically, the optimization of loss was considered as the balance between efficient absorption of the pump and a steep index dispersion is needed to maximize the performance of a material. In fact, the performance of multiple materials is evaluated using the proposed FOM, and all achieved their peak performance for moderate mobility ranges in the range of $20\ (cm^2/V-s)$, corresponding to $\varepsilon'' \sim 0.5$. This demonstrates that unlike other applications in plasmonics, pursuing exceptionally low loss ENZ materials is not necessary to maximize the IDRI effect, and can, in fact, be detrimental to the performance. However, this should be balanced with other factors, such as the need for efficient transmission of the probe, based on a given application. Ultimately, ENZ materials represent a unique platform for nonlinear interactions that is ultrathin and can be realized without complex nanofabrication while making use of intraband nonlinearities that are slow enough to enable a good temporal overlap with ultrafast laser systems yet fast enough for many ultrafast optical applications. By enabling a higher degree of predictive power and a platform for evaluating and optimizing new materials, practical devices for ultrafast, compact applications may move closer to reality.

## Appendix A: Energy Absorbed in a Thin Sample

The absorbed energy density can be calculated simply as the difference between the energy density just inside the sample and after some distance $dz$:

$$dU_{abs} = U(0) - U(dz) = \left(1 - R\right)\frac{I}{v_g}[1 - e^{-\alpha dz}] \tag{25}$$

where $I$ is the intensity of the pump beam external to the sample, $R$ is the reflection of the sample at the pump wavelength, and $v_g$ is the group velocity at the pump wavelength and accounts for slow light effects. We can shift to the time domain by substituting $dz = v_g dt$:

$$dU_{abs} = U(0) - U(dt) = \frac{I_i}{v_g}[1 - e^{-\alpha v_g dt}] \qquad (26)$$

where $I_i = I(1-R)$. We can simplify this by taking a Taylor expansion of the exponential function, keeping only the first-order term $1 - e^{-\alpha v_g dt} \approx \alpha v_g dt$, where $\alpha$ is measured absorption including slow light effects $\alpha = \alpha_0 n_g$. This approximation is valid in the thin sample regime.

$$dU_{abs} = \frac{I_i}{v_g}\alpha_0 n_g v_g dt \qquad (27)$$

The peak in the absorbed energy density ($\delta U_{abs}$) is then found by integrating over a time $\tau$ noting that there is a competing relaxation process of electron scattering ($e^{-t/\tau_{el}}$) which is removing energy from the electron gas. As an approximation, we may consider two cases, when $\tau_p \gg \tau_{el}$ and when $\tau_{el} \gg \tau_p$. In the first case, the electron relaxation time limits the accumulation of energy such that $\tau = \tau_{el}$ while in the latter case, the electrons accumulate energy for the entire duration of the pulse and $\tau = \tau_p$. The peak absorbed energy density is then:

$$\delta U_{abs} = I_i \alpha_0 n_g \tau = \frac{\alpha_0 n' n_g \tau}{2\eta_0} F_i^2 \qquad (28)$$

where $\tau$ is selected based on the criteria above.

In the case of Eq. 22, absorption is calculated from the Transfer Matrix Method and $(1-R)[1 - e^{-\alpha z}] \approx A$ such that $\delta U_{abs} = AI\tau / d$. The electron scattering rate is calculated through the optical mobility and electron effective mass as $\Gamma = q / (\mu_r m^*_{avg}(I = 0))$. Here the optical mobility is the value as determined through optical measurements, such as ellipsometry, as opposed to the Hall effect. Depending upon the material's structural quality the optical mobility can differ from the Hall mobility by as much as an order of magnitude [41], [59].

**Appendix B: Band Structure calculations**

The conduction band of zinc oxide [39] is taken as the basis for band calculations. To calculate effective mass and density of states, the band is fit by a hyperbola for simplicity. The hyperbola is defined by the parameters $a$ and $b$ such that $(E-a)^2/a^2 - k^2/b^2 = 1$ which sets the conduction band minimum, $E = 0$, as the minimum of the hyperbola. This leads to an E-k relationship of $E = \sqrt{a^2 + a^2 k^2/b^2} - a$ rather than the conventional parabolic band $E = \hbar^2 k^2/(2m^*)$. The parameters $a = 1.35$ $(eV)$ and $b = 0.211$ $(Å^{-1})$ are found to fit well to the AZO band structure in the range of $0 \leq E < 2.5eV$. From these constants, the effective mass and density of states can be calculated:

$$m^*(E) = \left[ \frac{2}{3} \frac{a^2}{\hbar^2 b^2 (E+a)} + \frac{1}{3} \frac{a^4}{\hbar^2 b^2 (E+a)^3} \right]^{-1} \quad (29)$$

$$DOS(E) = \frac{b^3}{\pi^2 a^3} \sqrt{(E+a)^2 - a^2} \, (E+a) \quad (30)$$

For comparison of materials, this hyperbolic approximation can be taken for many doped semiconductors.

Table 2 Hyperbolic Fit constants used in Section 5

| Material | $a$ parameter [eV] | $b$ parameter [Å⁻¹] | Approximate fit range [eV] |
|----------|--------------------|--------------------|----------------------------|
| Doped ZnO | 1.35 | 0.211 | 0-2.5 |
| ITO | 1.22 | 0.204 | 0-2 |
| CdO | 1.0 | 0.170 | 0-3 |
| GaN | 1.0 | 0.140 | 0-2.5 |
| Ga$_2$O$_3$ | 1.13 | 0.201 | 0-3 |

**Appendix C: Electron Relaxation**

A correction factor is needed to calculate the maximum absorbed energy in the case when the electron relaxation rate is comparable to the pulse width, $\tau_p \sim \tau_{el}$. Since the electron scattering times are on the order 10s of $(fs)$, the recovery time for the nonlinear effect in AZO was measured to be approximately 170 $(fs)$ and is comparable to the pump pulse width, 100 $(fs)$. This limits the peak temperature reached by the electron gas. In this scenario, we follow a similar logic as shown in Appendix A, where the pump pulse is now taken as a Gaussian in time with pulse width $\tau_p$, and our integration time is $\tau = \tau_p$. However, we need a correction factor to account for the electron relaxation within the pulse envelope, defined as:

$$\gamma(t) = \begin{cases} 1 & t < 0 \\ e^{-t/\tau_{el}} & t \geq 0 \end{cases}.$$ (31)

This model for relaxation provides us with a correction factor of $\eta = \max\left[\left(\gamma(t)/2\right)\left(1 + \mathrm{erf}\left(t/\tau_p\right)\right)\right]$, that is the ratio of the peak absorbed divided by the peak energy absorbed in the case of $\tau_{el} \sim \infty$. This factor is then applied to the pump intensity to find the maximum nonlinear response, $\delta U_{abs} = I_i \eta \alpha_0 n_g \tau_p$ under the thin sample approximation.

**Appendix D: Figure of Merit Calculation**

To compare various materials' nonlinear response, the permittivity of each material must go to zero at the same frequency. For each material, the background permittivity and effective mass are different, thus, the carrier concentration required to achieve ENZ at $\approx 1.3 \ (\mu m)$ changes as well. At significantly low mobilities (such as below $\mu_e \approx 20 \ (cm^2/V-s)$), $\Gamma$ approaches the order of $\omega$, shifting the ENZ wavelength, $\varepsilon_\infty = \dfrac{Nq^2}{\varepsilon_0 m^* m_0}\left(\omega_{ENZ}^2 + \Gamma^2\right)^{-1}$.

To keep the ENZ wavelength constant, the number of carriers, $N$, must compensate; this increases the average effective mass resulting in a significant decrease in FOM at low mobilities. For simplicity, the values for each of these parameters given a high mobility are shown in Table 1, while the values for calculation include this shift.

**7. Funding**

R.S. and N.K acknowledge funding from AFOSR grant FA9550-1-18-0151 and J.K. acknowledges DARPA NLM program HR00111820063.